\documentclass[journal]{IEEEtran}
\ifCLASSINFOpdf
\else
\fi
\usepackage{amsmath}
\usepackage{array}
\usepackage{booktabs}
\usepackage{amsbsy}

\usepackage{threeparttable}
\usepackage{dcolumn}
\usepackage{url}
\usepackage{cite}
\usepackage{psfrag}
\usepackage{amsfonts,amssymb}
\usepackage{epstopdf}
\usepackage{array}
\usepackage{multirow}
\usepackage{bm}
\usepackage{subfigure}
\usepackage{verbatim}
\usepackage{colortbl}
\usepackage{stfloats}
\usepackage{bm}
\usepackage{algorithm}

\usepackage{bigstrut}
\usepackage{amsmath}
\usepackage{mathtools}
\usepackage{algpseudocode}
\usepackage{float}
\usepackage{epsfig}
\usepackage[justification=centering]{caption}

\begin{document}
\newtheorem{theorem}{Theorem}
\newtheorem{lemma}{Lemma}
\newtheorem{definition}{Definition}
\title{ATSFFT: A Novel Sparse Fast Fourier Transform Enabled With Sparsity Detection}

\author{Sheng Shi, Runkai Yang, Xinfeng Zhang
        and Haihang You
\thanks{S. Shi, R. Yang and H. You are with Institute of Computing Technology, Chinese Academy of Sciences, Beijing, 100190, e-mail:(shisheng, yangrunkai, youhaihang)@ict.ac.cn}
\thanks{X. Zhang is with University of Chinese Academy of Sciences, Beijing, 100049,
e-mail: xfzhang@ucas.ac.cn}
}

\maketitle

\begin{abstract}
The Fast Fourier Transform(FFT) is a classic signal processing algorithm that is utilized in a wide range of applications. For image processing, FFT computes on every pixel's value of an image, regardless of their properties in frequency domain.
The Sparse Fast Fourier Transform (SFFT) is an innovative algorithm for discrete Fourier transforms on signals that possess characteristics of the sparsity in frequency domain. A reference implementation of the algorithm has been proven to be efficient than modern FFT library in cases of sufficient sparsity. However, the SFFT implementation has a critical drawback that it only works reliably for very specific input parameters, especially signal sparsity $k$, which hinders the extensive application of SFFT. In this paper, we propose an Adaptive Tuning Sparse Fast Fourier Transform (ATSFFT), which is a novel sparse fast fourier transform enabled with sparsity detection. In the case of unknown sparsity $k$, ATSFFT is capable of probing the sparsity $k$ via adaptive dynamic tuning and completing the sparse Fourier transform. Experimental results show that ATSFFT outperforms SFFT while it is able to control the computation error better than SFFT. Furthermore, ATSFFT achieves an order of magnitude of performance improvement than the state-of-the-art FFT library, FFTW.
\end{abstract}


\IEEEpeerreviewmaketitle

\section{Introduction}
Nowadays, the development of information technology has reached unprecedent level. The fast growing computing power stimulate emerging technologies that promote the development of human society. Magnetic Resonance Imaging (MRI)\cite{jour0}, Light Field Photography \cite{jour01}, Radio Astronomy \cite{jour02} and etc. are the applications of image processing that have wide impact on health care, technology and science. There is tremendous demand of high efficient signal processing techniques for drasticlly increasing amount of images to process. The discrete Fourier transform (DFT) is one of the most fundamental and important numerical algorithms which plays the central role in signal processing\cite{jour03}\cite{jour04}, communications, and audio/image/video compression \cite{jour1}. The Fast Fourier Transform (FFT) \cite{jour2} that computes the DFT of an $n$-size signal in $O(n\log n)$ time greatly simplifies the complexity of the DFT and boosts the performance substantially, thus is utilized by a broad range of applications.

The general algorithms for computing the exact DFT necessitate the time that is at least proportional to its size $n$. However, it is well known that most image signals posses sparsity in frequency domain. That is, the image signals have naturally sparse representations with respect to fixed Fourier basis. This property has been widely used in various applications including High Efficiency Video Coding (HEVC)\cite{jour3}\cite{jour4}\cite{jour5}\cite{book1}, compressed sensing \cite{book2}\cite{jour6}\cite{book3} and radio astronomy[reference]. Therefore, for sparse image signals, the lower bound $\Omega(n)$ of the DFT complexity no longer applies. It is crucial to study the new strategy of the Fourier transform based on image sparsity. In 2012, Hassanieh \emph{et al} proposed one-dimensional sparse fast Fourier transform \cite{book4} \cite{book5} which is faster than traditional FFT, needless to say, the algorithm demonstrates a promising approach.

However, the SFFT implementation has the drawback that it only works reliably for very specific input parameters, especially signal sparsity $k$. This drawback hinders the extensive applications of SFFT.
In addition, two-dimensional sparse Fourier transform can not simply be implemented by utilizing two separate one-dimensional Sparse Fourier transform. Since two-dimensional transform for image signals are more widely used in practival applications, we propose a new two-dimensional Fourier transform that takes advantage of images sparsity(2D-SFFT) \cite{book6}\cite{book7}. Furthermore, we propose an Adaptive Tuning Sparse Fast Fourier Transform (ATSFFT) to improve the accuracy and robustness of SFFT. With adaptive tuning, ATSFFT is able to probe the sparsity $k$ automatically and obtain the Fourier coefficients of signals. Experimental results show that ATSFFT not only can control the error better than SFFT, it could outperform SFFT with an order of magnitude of speedup in some cases.

The remainder of this paper is organized as follows. Section \ref{sec:format} presents details of the proposed algorithm of two-dimensional Sparse Fourier Transform (SFFT). Section \ref{sec:pagestyle} describes the Adaptive Tuning Sparse Fast Fourier Transform (ATSFFT) algorithm. Experimental results are shown in Section \ref{sec:typestyle}, and  at last we conclude the paper in Section \ref{sec:majhead}.

\section{Two-dimensional Sparse Fast Fourier Transform}
\label{sec:format}
First, we lay out several conventions and notations that are used in this paper. A space-domain image is represented as a tow-dimensional matrix $x\in{C^{N\times N}}$, the Fourier spectrum of the image is represented as $\hat{x}$. We assume that $N$ is a power of 2, the notation $[N]$ is defined as the set $\{0,1,...,N-1\}$, and $[N]\times[N]={[N]}^2$ denotes the $N\times N$ grid $\{(i,j):i\in[N],j\in[N]\}$. The image support is denoted by $supp(x)\subseteq [N]\times[N]$. 
All matrix indices are the calculated modulo of the matrix size, e.g. $x_{i,j}$ of image $x$ is actually $x_{i~\mathbf{mod}~n,j~\mathbf{mod}~n}$.
A set of matrix elements can be written as a matrix subscripted with a set of indices, for example $x_{I,J}=\{x_{i,j}|i\in I,j\in J\}$. In addition, we assume that $B$ is a power of 2, and $N$ can be divisible by $B$. 

We define $\omega=e^{-2\pi{i}/N}$ to be a primitive $N$-th root of unity. In the following sections, we will use the following definition of the 2D-DFT without the constant scaling factor:
\begin{eqnarray}
\nonumber&\hat x_{i,j}=\sum_{u\in[N]}\sum_{v\in[N]}x_{u,v}\omega^{iu+jv}, ~~(i,j)\in{\Omega_N}\\
&\Omega_N=\{(i,j)|0\leq i\leq {N-1},0\leq j\leq {N-1}\}
\end{eqnarray}
This makes some of the proof easier, but it is not considered relevant in practical implementations.

\subsection{Hash Function}
The 2D-SFFT algorithm firstly constructs and utilizes a hash function to extract useful information of an image. The hash function consists of random spectrum permutation, filtering and subsampling in frequency domain.
\subsubsection{Random Spectrum Permutation}
Normally, we do not have access to the input images' Fourier spectrum since it would involve performing DFT. The spectrum permutation is the primary component of the 2D-SFFT, which is defined in Definition~\ref{def:perm}. It aims to tear apart the nearby coefficients to reorder the image's frequency-domain $\hat x$:

\begin{definition}\label{def:perm}
Let $\sigma_1$ and $\sigma_2$ be invertible modulo $n$, i.e. $\gcd(\sigma_1,n)=1$, $\gcd(\sigma_2,n)=1$, and $\tau_1\in[n]$, $\tau_2\in[n]$. Then, $i\rightarrow\sigma_1 i+\tau_1~\mathrm{mod}~n$ and $j\rightarrow\sigma_2 j+\tau_2~\mathrm{mod}~n$ are permutations on $[n]$. The associated permutation $P_{\sigma_1,\sigma_2,\tau_1,\tau_2}$ on a matrix x is then given by
\begin{equation}
(P_{\sigma_1,\sigma_2,\tau_1,\tau_2}x)_{i,j}=x_{\sigma_1i+\tau_1,\sigma_2j+\tau_2}
\end{equation}
\end{definition}
When a permutation is applied to an image $x$ in space domain, the image's frequency domain $\hat x$ is also permuted. This interesting property is derived in Lemma~\ref{lemma:perm}.

\begin{lemma}\label{lemma:perm}
Let $P_{\sigma_1,\sigma_2,\tau_1,\tau_2}$ be a permutation and $x$ be an two-dimensional vector. Then
\begin{equation}
\widehat{(P_{\sigma_1,\sigma_2,\tau_1,\tau_2}x)}_{{\sigma_1}i,{\sigma_2}j}=\hat x_{i,j}\omega^{-({\tau_1}i+{\tau_2}j)}, (i,j)\in{\Omega_n}
\end{equation}
\end{lemma}
$Proof.$ For $(i,j)\in{\Omega_n}$,
\begin{equation}
\widehat{(P_{\sigma_1,\sigma_2,\tau_1,\tau_2}x)}_{i,j}=\sum_{u\in[n]}\sum_{v\in[n]}x_{{\sigma_1u+\tau_1},{\sigma_2v+\tau_2}}\omega^{iu+jv}
\end{equation}
with $a_1=\sigma_1u+\tau_1$, $a_2=\sigma_2v+\tau_2$
\begin{eqnarray}
\nonumber&\widehat{(P_{\sigma_1,\sigma_2,\tau_1,\tau_2}x)}_{i,j}\\
\nonumber&=\sum\limits_{a_1\in[n]}\sum\limits_{a_2\in[n]}x_{a_1,a_2}
\omega^{\frac{(a_1-\tau_1)}{\sigma_1}i+\frac{(a_2-\tau_2)}{\sigma_2}j}\\
\nonumber&=\omega^{-(\frac{\tau_1}{\sigma_1}i+\frac{\tau_2}{\sigma_2}j)}\sum\limits_{a_1\in[n]}\sum\limits_{a_2\in[n]}x_{a_1,a_2} \omega^{(\frac{\tau_1}{\sigma_1}i+\frac{\tau_2}{\sigma_2}j)}\\
&=\omega^{-(\tau_1{\sigma_1}^{-1}i+\tau_2{\sigma_2}^{-1}j)}\hat{x}_{{{{\sigma_1}^{-1}}i,{{\sigma_2}^{-1}}j}}
\end{eqnarray}
The Lemma follows by substituting $i={\sigma_1}i$, $j={\sigma_2}j$. Note that $\omega^{-({\tau_1}i+{\tau_2}j)}$ changes the phase, but does not change the magnitude of $\hat x_{i,j}$.

The permutation in the 2D-SFFT algorithm allows to permute the image's Fourier spectrum by modifying the image's space-domain $x$.

\subsubsection{Window Function}

In order to achieve substantial performance improvement, the 2D-SFFT only uses partial input image for computation. The standard window function acts like a filter, it supplies the sparse Fourier transform algorithm with a subset of the Fourier coefficients. Ideally, however, we would like the pass region of the filter to be as flat as possible to avoid spectral leakage. Specifically, two-dimensional flat Guassian window functions are used in 2D-SFFT.

\begin{figure}
  \centering
  \includegraphics[width=0.3\textwidth]{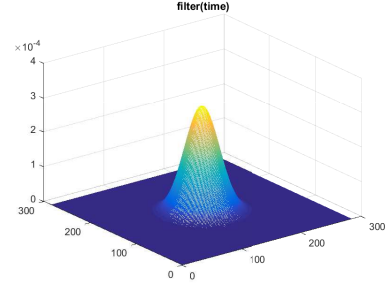}
  \caption{2D Gaussian window function}
  \label{fig:1}
\end{figure}

The two-dimensional flat Guassian window function can be obtained from a 2D Gaussian standard window function which is shown in Figure~\ref{fig:1} by convolving it with a two-dimensional "box car" window function which can be presented as:
\begin{eqnarray}
r(x,y)=
\left\{
\begin{array}{lll}
1, & (x,y)\in D\\
0, & (x,y)\in D^{'}
\end{array}
\right.
\end{eqnarray}
where $D=\{(x,y)|-\frac{b}{2}\leq x \leq \frac{b}{2}, -\frac{b}{2}\leq y \leq \frac{b}{2}\}$.

\begin{figure}
\begin{minipage}{0.49\linewidth}
  \centerline{\includegraphics[width=1\textwidth]{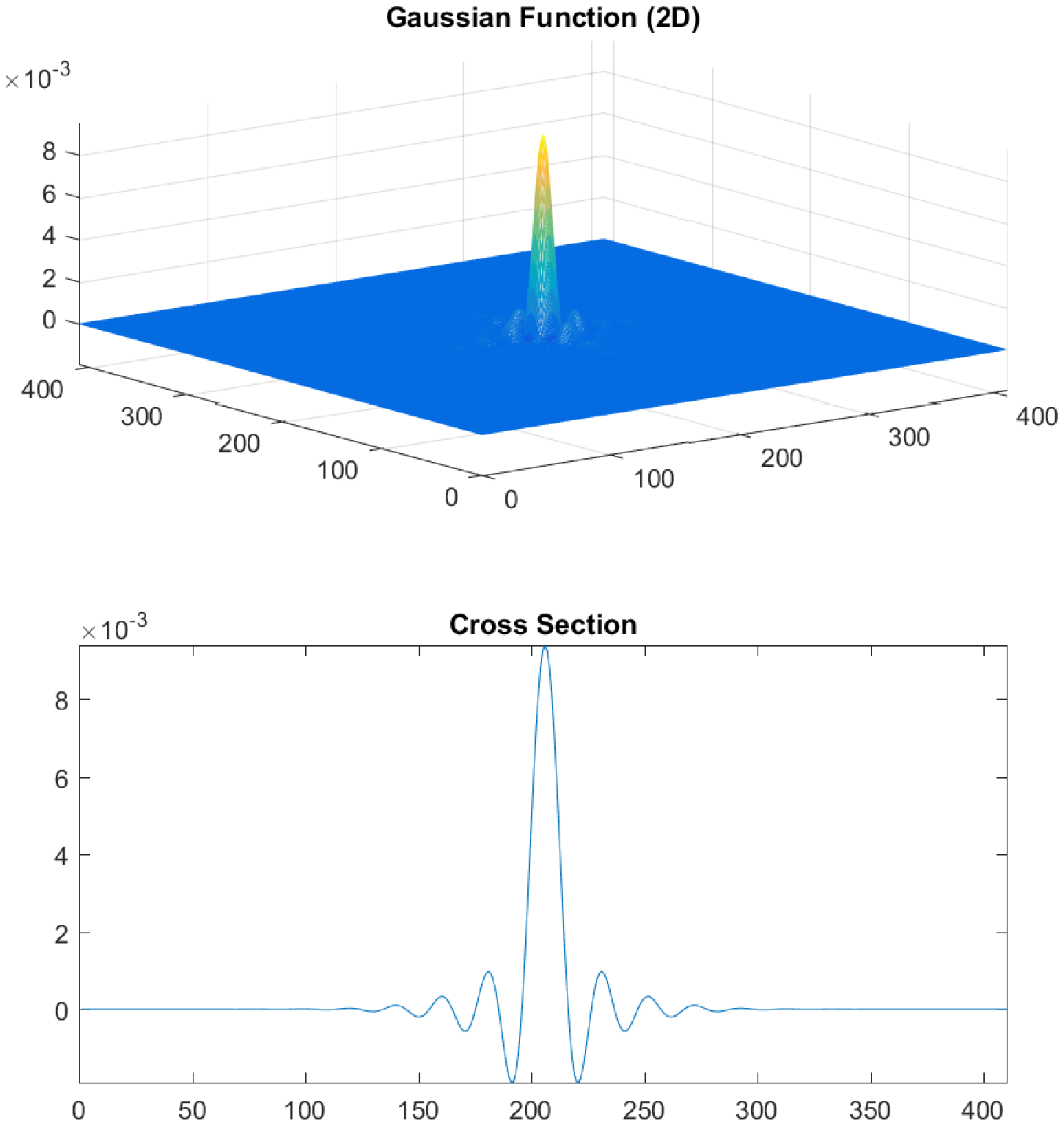}}
  \centerline{\scriptsize{(a)}}
  \centerline{}
\end{minipage}
\hfill
\begin{minipage}{0.49\linewidth}
  \centerline{\includegraphics[width=1\textwidth]{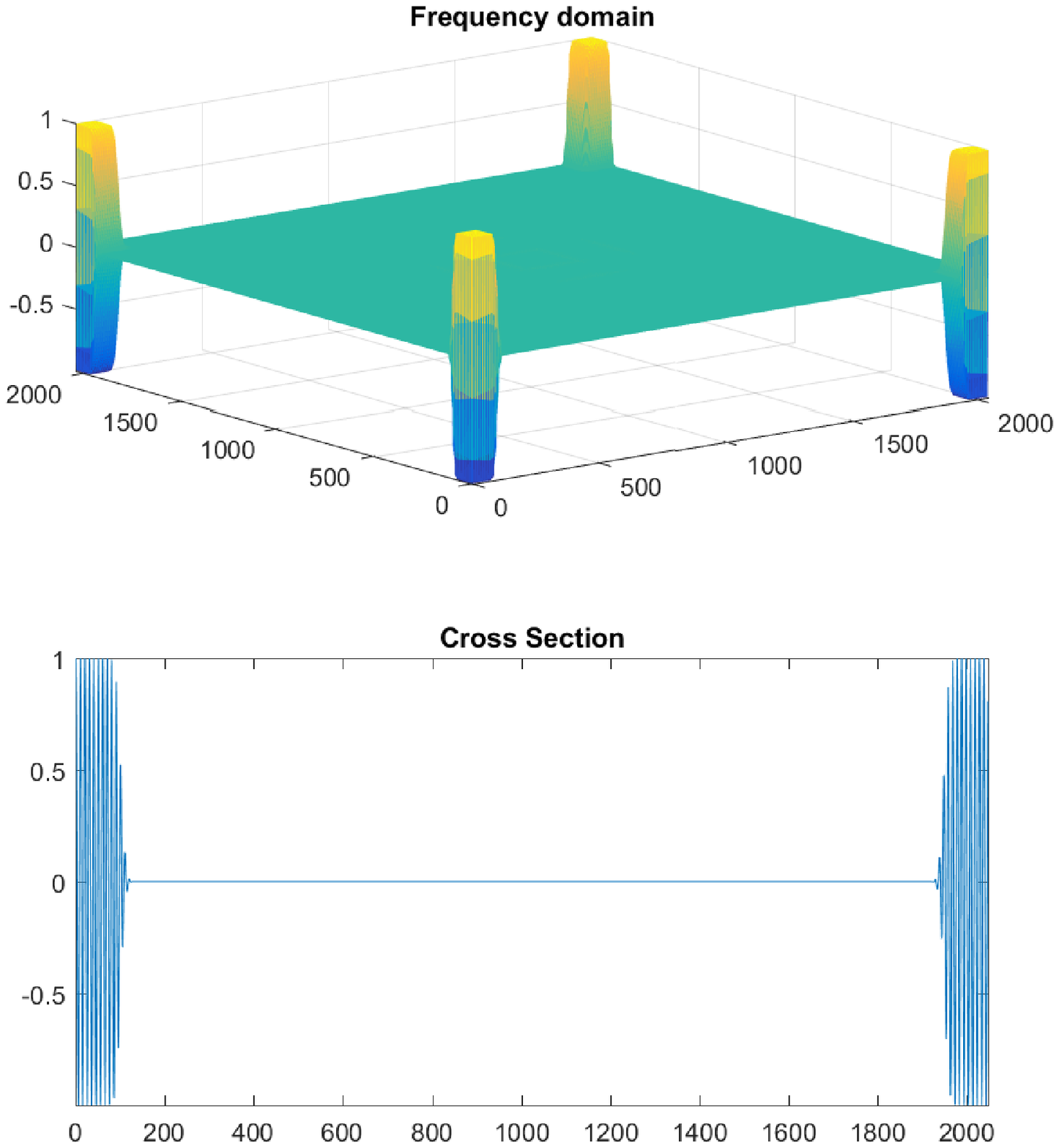}}
  \centerline{\scriptsize{(b)}}
  \centerline{}
\end{minipage}
\caption{2D Gaussian flat window function in time domain and frequency domain (a) 2D Gaussian Function $G$ (b) 2D Gaussian Function $\hat{G}$}
\label{fig:2}
\end{figure}

The 2D Gaussian window function is defined as

\begin{equation}
f(x,y)=A exp [-({\frac{(x-x_0)^2}{2{\sigma}^2_x}}+{\frac{(y-y_0)^2}{2{\sigma}^2_y}})]
\end{equation}
By applying convolution of (8) and (9), we have the 2D Gaussian flat window function $G$. Figure~\ref{fig:2} shows the function in time domain and frequency domain.

Using 2D Gaussian flat window function $G$, part of size $|supp(G)|$ can be extracted out from $P_{\sigma_1,\sigma_2,\tau_1,\tau_2}x$ by multiplying $G$ with $x$ and neglecting the coefficients with value of zero. According to the convolution theorem, the multiplication is equivalent to a convolution of $\hat{G}$ and $\hat{x}$. The filtering process can expand the area of non-zero coefficients. This step is to prepare for the subsequent sub-sampling and reverse steps, and further increase the probability of detection of non-zero coefficients.

\subsubsection{Fast Subsampling and DFT}
\begin{lemma}
Let $B\in N$ divide $n$, $x$ be an $N\times N$ two-dimensional matrix and $y$ be a $B\times B$ two-dimensional matrix with $y_{i,j}=\sum_{u\in[n]}\sum_{v\in[n]}x_{i+Bu,j+Bv}$ for $i=1,...,B$,~$j=1,...,B$. Then, $\hat y_{i,j}=\hat x_{i(n/B),j(n/B)}$
\end{lemma}

Lemma 2 effectively reduces dimension by subsampling image in time domain and summing up the result. Since image is sparse in frequency domain, dimension reduction can reduce the complexity of position searching and amplitudes of non-zero elements.

\begin{figure*}
\begin{minipage}{0.24\linewidth}
  \centerline{\includegraphics[width=1\textwidth]{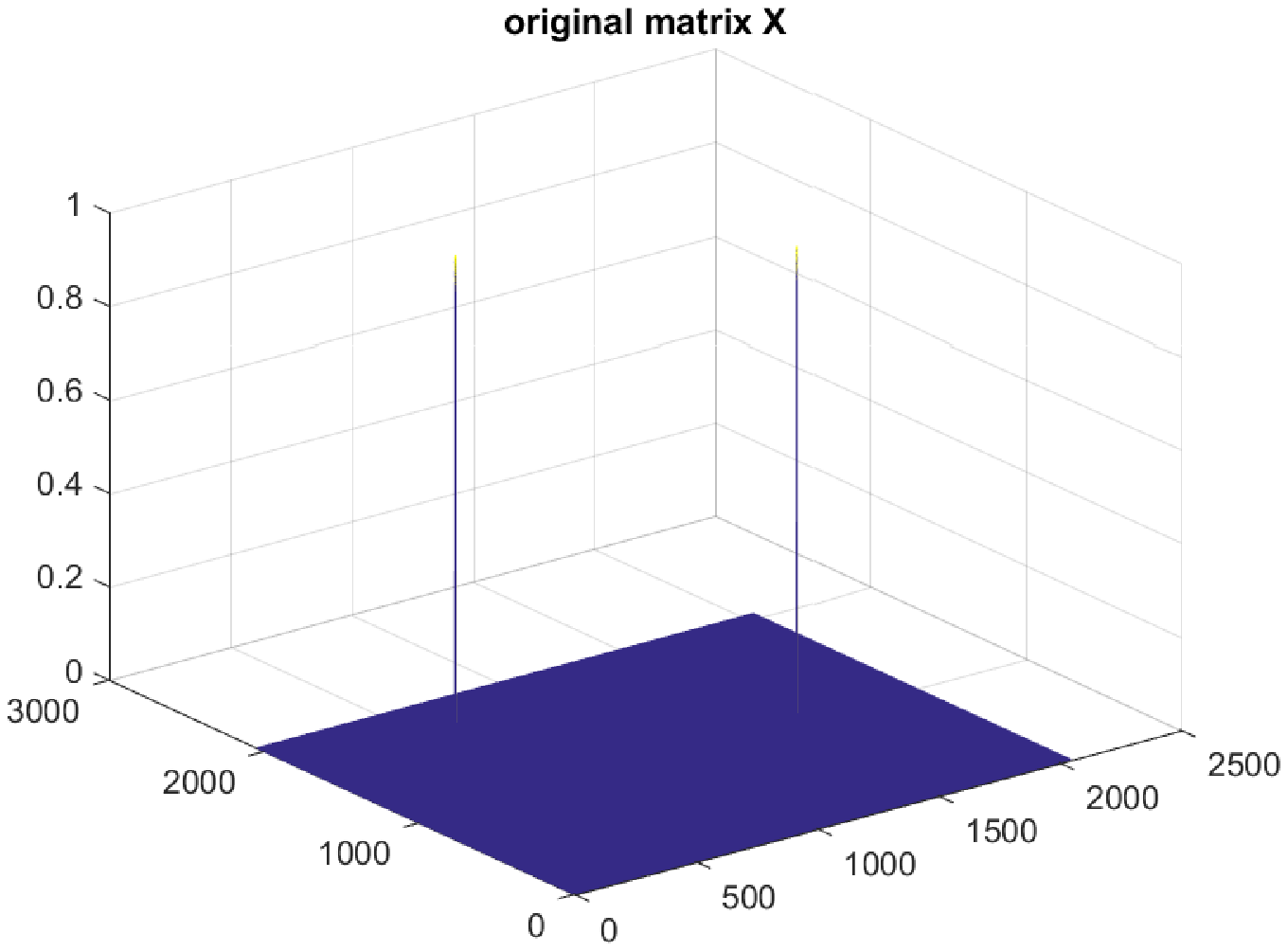}}
  \centerline{\scriptsize{(a)}}
  \centerline{}
\end{minipage}
\hfill
\begin{minipage}{0.24\linewidth}
  \centerline{\includegraphics[width=1\textwidth]{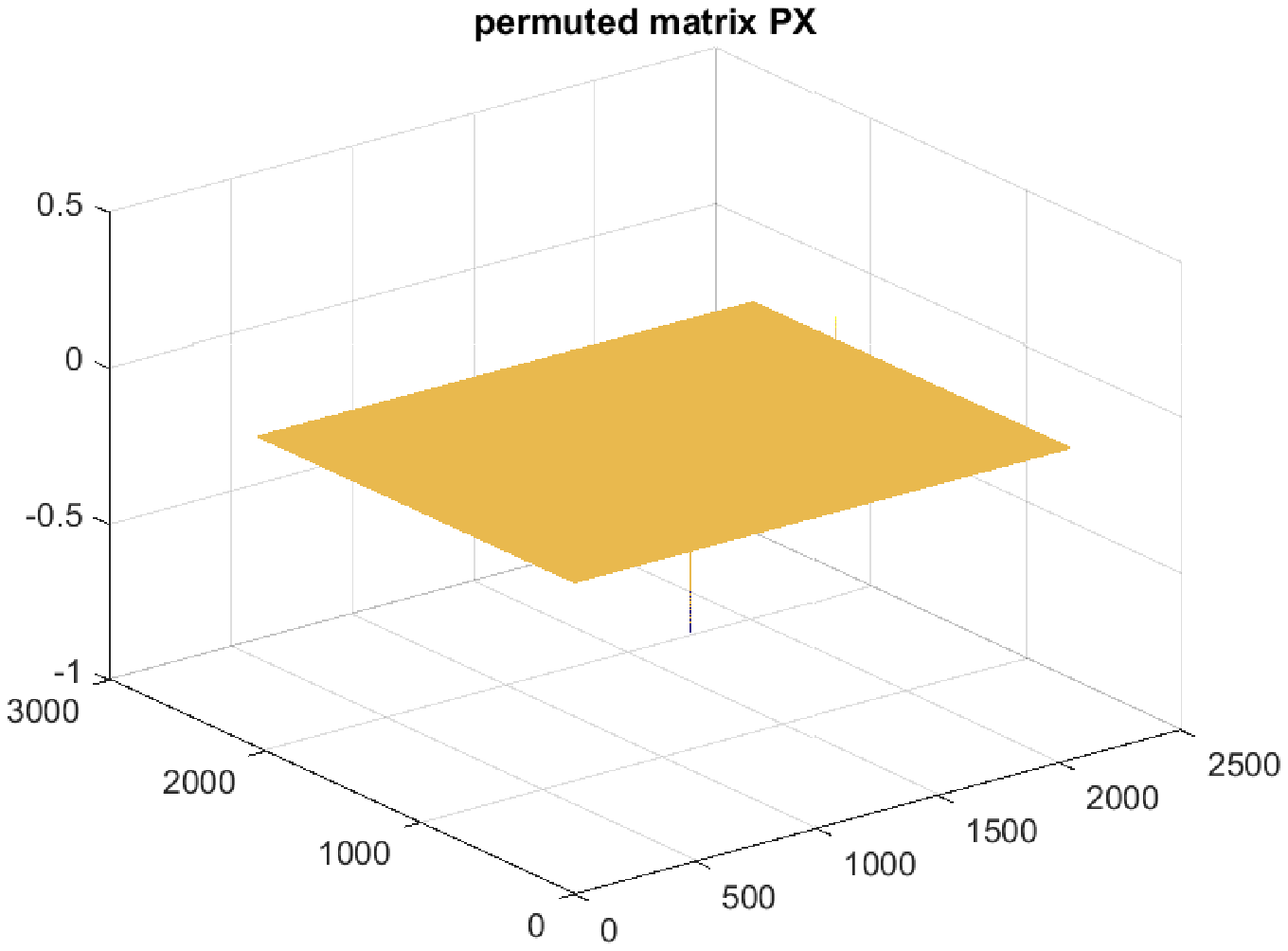}}
  \centerline{\scriptsize{(b)}}
  \centerline{}
\end{minipage}
\hfill
\begin{minipage}{0.24\linewidth}
  \centerline{\includegraphics[width=1\textwidth]{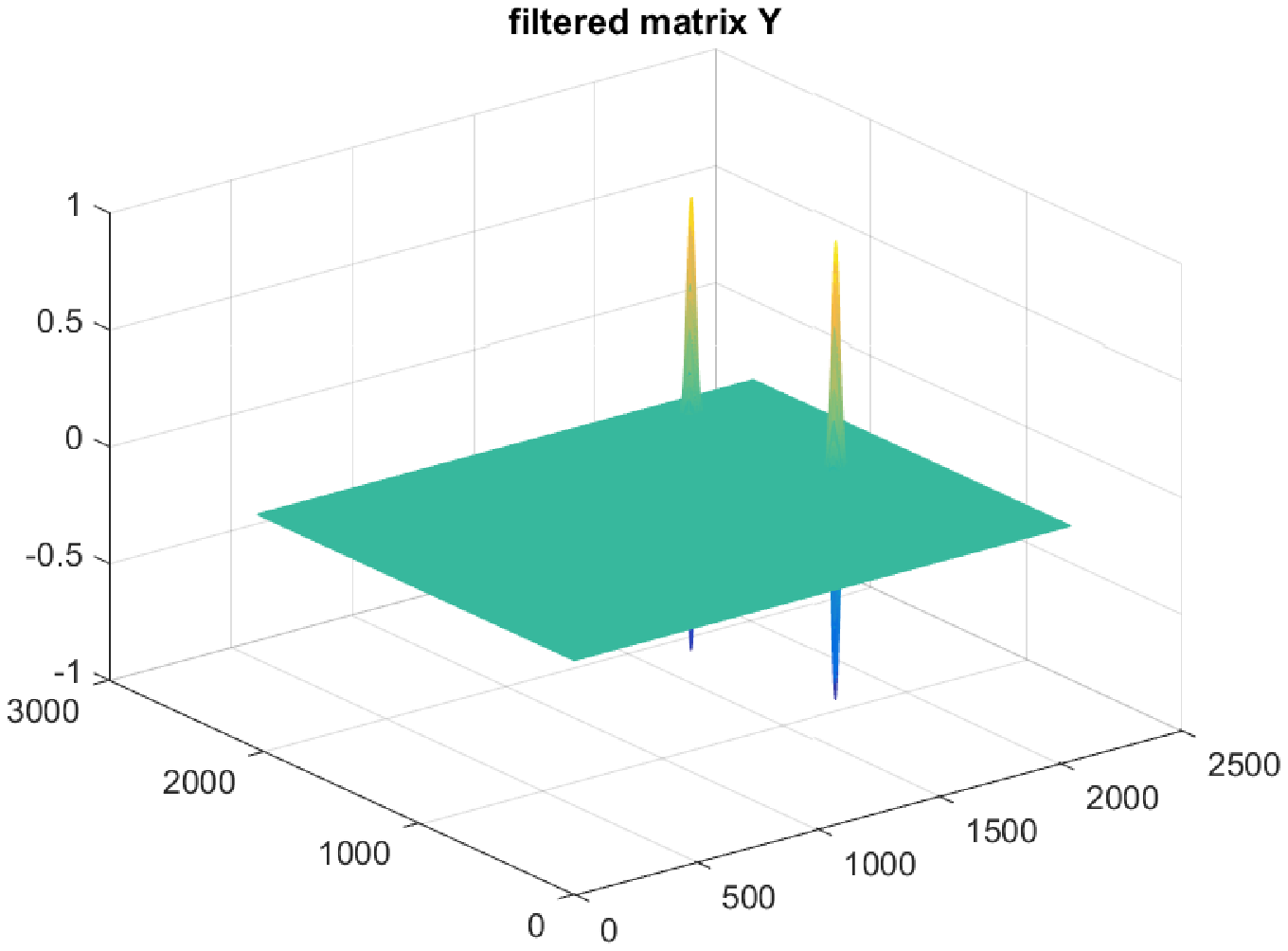}}
  \centerline{\scriptsize{(c)}}
  \centerline{}
\end{minipage}
\hfill
\begin{minipage}{0.24\linewidth}
  \centerline{\includegraphics[width=1\textwidth]{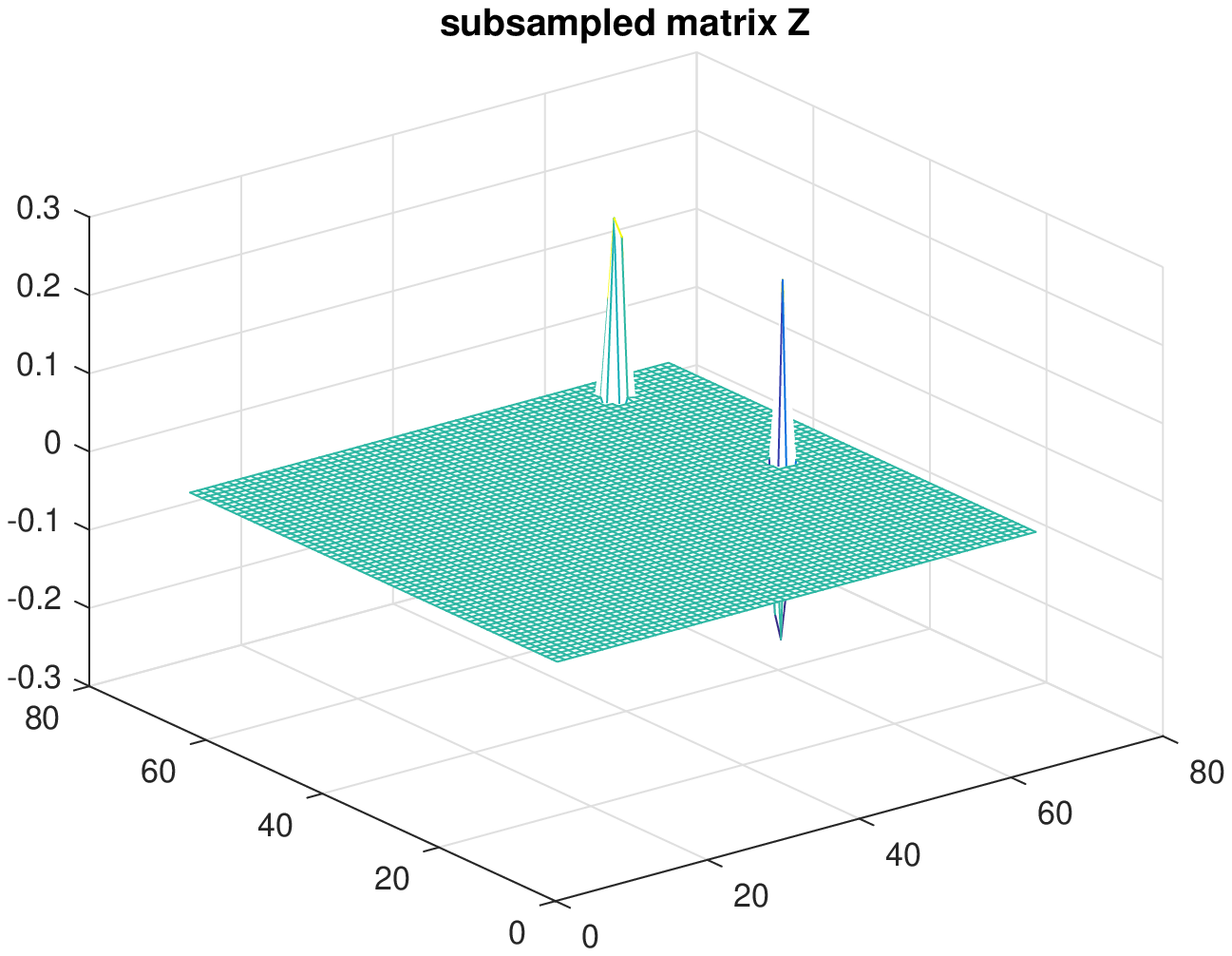}}
  \centerline{\scriptsize{(d)}}
  \centerline{}
\end{minipage}
\caption{The process of permutation, filtering and subsampling on image $(k=2)$ (a) Original image $\hat{x}(k=2)$,(b) permuted image $\widehat{(P_{\sigma_1,\sigma_2,\tau_1,\tau_2}}x)$,(c) filtered image $\widehat{(G\cdot P_{\sigma_1,\sigma_2,\tau_1,\tau_2}x)}$,(d) Subsampled image in frequency domain}
\label{fig:3}
\end{figure*}

An example of an image in frequency domain with sparsity $(k=2)$ is show in Figure~\ref{fig:3}(a). The process of random spectrum permutation, filtering and subsampling are shown in Figure~\ref{fig:3}. Permutation can separate nearby coefficients so that the non-zero coefficients can be approximately uniform distributed. Filtering process can expand the area of non-zero coefficients to increase the detection probability. Subsampling effectively reduces complexity.

Random spectrum permutation, filtering and subsampling define the hash function $h_{\sigma_1,\sigma_2}: [N\times N] \to [B\times B]$
\begin{equation}
h_{\sigma_1,\sigma_2}(i,j)=round ({\sigma_1}{\sigma_2}ij\frac{B^2}{N^2}), i\in[N], j\in[N]
\end{equation}
Hash function $h_{\sigma_1,\sigma_2}$ maps each of the $N\times N$ coordinates of the input image to one of $B\times B$ bins.

\subsection{Two-dimensional Sparse Fourier Transform Algorithm}
\label{sec:majhead}
The 2D-SFFT consists multiple executions of two kinds of operations: seek location and estimate coefficient. The seek location operation is to generate a list of candidate coordinates which have certain probability of being indices of non-zero coefficients in frequency domain. While the estimate coefficient operation is used to exactly determine the frequency coefficients. The implementation of the estimate coefficient also uses hash function. Given a set of coordinates $I$, $\hat{x}_{i,j}$ can be estimated by
\begin{equation}
\hat x^{'}_{i,j}={\hat Z_{h_{\sigma_1,\sigma_2}(i,j)}\omega^{{\tau_1}i+{\tau_2}j}}/\hat{G}_{o_{\sigma_1,\sigma_2}(i,j)}
\end{equation}
which basically removes the phase change due to the permutation and the filtering.

\begin{figure}[t]
\centering
\begin{minipage}{0.8\linewidth}
  \centerline{\includegraphics[height=0.8\textwidth]{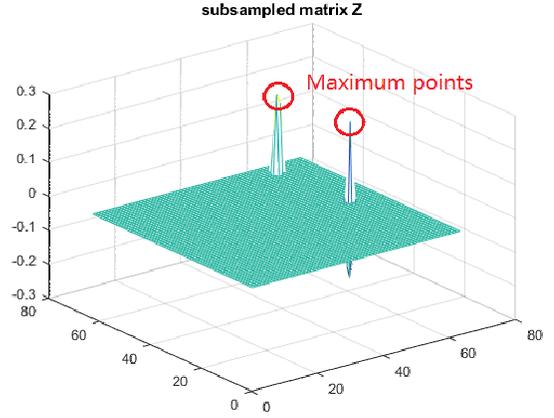}}
  \centerline{}
\end{minipage}
\caption{The relationship of the number of local maximum points and sparisty $k$ after hash function in frequency domain}
\label{fig:5}
\end{figure}

A simplified workflow diagram of 2D-SFFT is shown in Figure~\ref{fig:5}. After running multiple iterations of the seek location operation we only keep coordinates emerge in at least half of the seek location loops. For the coordinates $I^{'}$, the median of the corresponding outputs of the $L$ rounds of the estimate coefficient operation is set to be the frequency coefficient.
\begin{figure}
  \centering
  \includegraphics[width=0.45\textwidth]{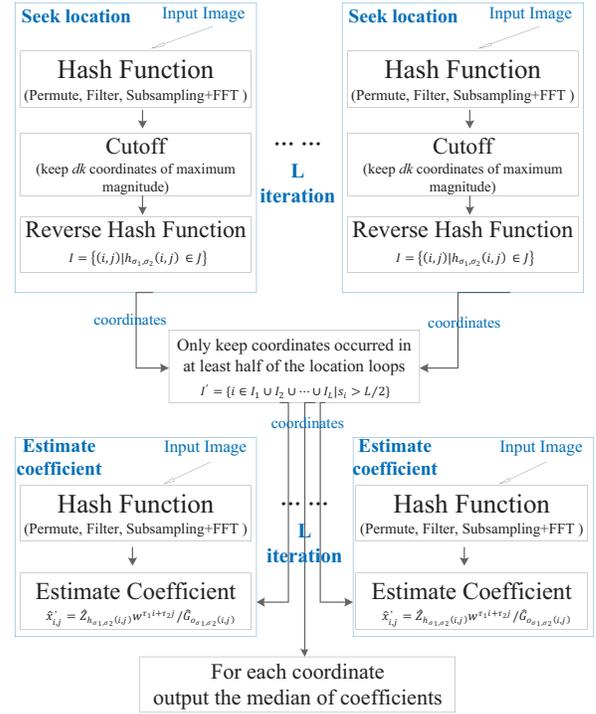}
  \caption{A simplified workflow diagram of 2D-SFT}
  \label{fig:4}
\end{figure}

\begin{algorithm}[h]
\caption{2D Sparse Fast Fourier Transform Algorithm}
\label{alg::conjugateGradient}
\begin{algorithmic}[1]
\Require
image size $N$;
image sparsity $k$;
iteration times $loops$;
$threshold$ and $tolerance$;
\Ensure
time and error of SFFT;
\State initial $sigmal$, $tau$, $num$, $x\_samp$ and etc;
\For{$i=0$; $i<loops$; $i++$}
\State do hash function: permute, filter and subsample;
\State do 2D-dft with length $B\times B$;
\State find $num$ largest indices in $x\_samp$;
\State do reverse hash function;
\EndFor
\State keep coordinates that occured in at least half of the location loops
\For{$i=0$; $i<loops$; $i++$}
\State do hash function: permute, filter and subsample;
\State estimate coefficient;
\EndFor
\State For each coordinate, ouput the median of coefficients from all estimation loops.
\end{algorithmic}
\end{algorithm}

\section{Adaptive Tuning Sparse Fast Fourier Transform}
\label{sec:pagestyle}
The drawback of the SFFT algorithm is that it only works reliably with prior knowledge of signal sparsity $k$. This drawback prevent it to be utilized by the wide range of applications. Although it is well known that most signals possess the characteristics of sparsity, it is almost impossible to foreknow the sparsity $k$ of the signals. Therefore, we propose a innovative Adaptive Tuning Sparse Fast Fourier Transform (ATSFFT) algorithm. By adaptive dynamic tuning, the ATSFFT can predict the sparsity $k$ and execute the Fourier transform. Experimental results show that the ATSFFT outperforms the SFFT, it also can have a better control of the error.

\subsection{Adaptive Tuning Iteration}
The basic idea behind hash function is to hash the $N\times N$ coefficients of the input image into a small number of $B\times B$ bins. In SFFT, $B$ is determined by $k$, which is set to $\sqrt{Nk}$. From these bins, SFFT selects and only keeps coordinates of top $dk$ points with the largest magnitudes. The actual locations of the Fourier coefficients in the frequency domain are approximated based on these coordinates. We find that after the execution of the hash function, the number of local maximum points is the same with the number of original image's non-zero coefficients in frequency domain. Specifically, the hash function hashes $2^{11}\times{2^{11}}$ points of the image shown in Figure~\ref{fig:3}(a) into $2^{6}\times{2^{6}}$ bins. The number of local maximum points is $2$ which is the same with the image sparsity($k=2$). Therefore, by finding the local maximum points, we can avoid prior knowledge sparsity $k$.

ATSFFT initially sets a small value to $B_0$ and $k_0$. After hashing, we can get $k_1$ by counting the number of local maximum values in $B\times B$ matrix. Next, $B_1$ can be updated by analyzing the relationship between $k_0$ and $k_1$. For the convenience of description, the change ratio factor $r$ of $k$ is defined as follows,

\begin{equation}
r_i=|{1-\frac{k_i}{k_{i-1}}}|.
\end{equation}
The iterative process is shown as follows

\begin{eqnarray}
B_{i+1}=
\left\{
\begin{array}{lll}
{\varepsilon_1}B_i, \quad\quad & 0{\leq}{r_i}<{\delta}_1\\
B_i, \quad\quad & {\delta}_1{\leq}{r_i}<{\delta}_2\\
(1+{\varepsilon_2})B_i, \quad\quad & {\delta}_2{\leq}{r_i}\\
\end{array}
\right.
\end{eqnarray}
where $0<{\delta_1}<{\delta_2}<1$ and $0<{\varepsilon_1}<{\varepsilon_2}<1$.
When $k_{i}$ approaches $k_{i-1}$ within a very small range of ${\delta}_1$, we consider $B_{i}$ adapts to the full scatter point set where non-zero elements can adequately distributed, so $B_{i+1}$ need to be decreased to find the better size. When $k_{i}$ approaches $k_{i-1}$ within a appropriate range between ${\delta}_1$ and ${\delta}_2$, we consider $B_{i}$ adapts to the appropriate scatter point set. If we further reduce the size of $B$, it will lead to large area overlap. Therefore, $B_{i+1}$ is kept the same with $B_{i}$. When $k_{i}$ fluctuates within a large range of $k_{i-1}$, we consider $B_{i}$ is too small to lead large area overlap, so we need to continue to increase the value of $B$. Since $B$ is a power of 2, and $N$ can be divisible by $B$. ATSFFT sets ${\varepsilon_1}=1/2$, ${\varepsilon_2}=1$, ${\delta_1}=2\%$, ${\delta_2}=5\%$. 

\begin{eqnarray}
B_{i+1}=
\left\{
\begin{array}{lll}
{1/2}B_i, \quad\quad & 0{\leq}{r_i}<{2\%}\\
B_i, \quad\quad & {2\%}{\leq}{r_i}<{5\%}\\
2B_i, \quad\quad & {5\%}{\leq}{r_i}\\
\end{array}
\right.
\end{eqnarray}


\begin{algorithm}[h]
\caption{Adaptive Tuning Sparse Fast Fourier Transform Algorithm}
\label{alg::conjugateGradient}
\begin{algorithmic}[1]
\Require
image size $N$;
iteration times $loops$;
$threshold$ and $tolerance$;
\Ensure
time and error of ATSFFT;
\State initial $sigmal$, $tau$, $num$, $x\_samp$, $delta$ and etc;
\State initial $B_0$, $B_1$ and $k_0$;
\For{$i=0$; $i<loops$; $i++$}
\State do hash function: permute, filter and subsample;
\State do 2D-dft with length $B\times B$;
\State find $num$ local maximum indices in $x\_samp$;
\State compute $ratio={k[i]}/{k[i-1]}$;
\If{${ratio>(1+delta2)}||{ratio<(1-delta2)}$}
\State $B[i+1]=B[i]\times{2}$;
\Else
\If{${ratio>(1+delta1)}||{ratio<(1-delta1)}$}
\State $B[i+1]=B[i]$;
\Else
\State $B[i+1]=B[i]/{2}$;
\EndIf
\EndIf
\State do reverse hash function;
\EndFor
\State keep coordinates occured exceed threshold;
\For{$i=0$; $i<loops$; $i++$}
\State do hash function: permute, filter and subsample;
\State estimate coefficient;
\EndFor
\State For each coordinate, ouput the median of coefficients from all estimation loops.
\end{algorithmic}
\end{algorithm}

\subsection{Adaptive Tuning Sparse Fast Fourier Transform Algorithm}
\begin{figure}
  \centering
  \includegraphics[width=0.45\textwidth]{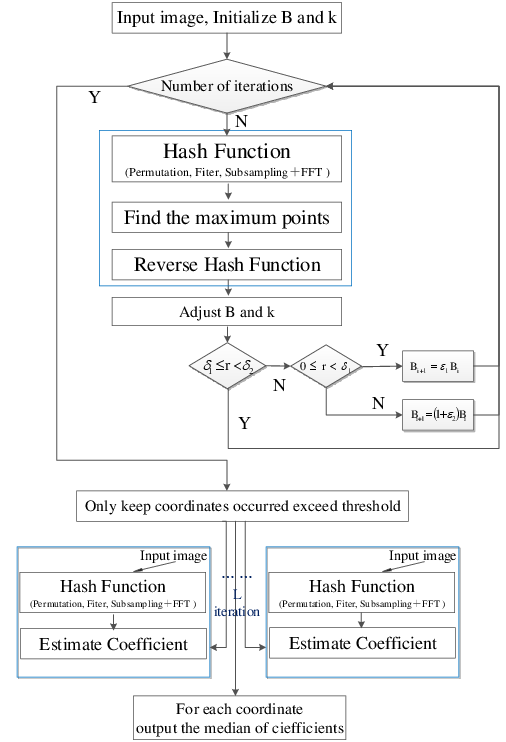}
  \caption{A simplified workflow diagram of ASFFT}
  \label{fig:7}
\end{figure}

A simplified workflow diagram of ATSFFT is shown in Figure~\ref{fig:7}. Similar to the SFFT, ATSFFT consists two kinds of operations: seek location and estimate coefficient. In the operation of seek location, by running multiple adaptive tuning iterations, it is able to identify the sparsity $k$ and locate candidate coordinates with a high probability of being one of the $k$ non-zero coordinates. Given the set of candidate coordinates , the ATSFFT can use coefficient estimation to precisely determine the frequency coefficients, which basically remove the phase change due to the permutation and the effect of the filtering.

\section{Numerical Experiments}
\label{sec:typestyle}
In this section, we present numerical experimental results that compare runtime, speedup and error of ATSFFT, SFFT and FFTW, which is a FFT algorithms implementation known to be one of the fastest FFT libraries \cite{book8}. For the experiments, $k$ frequencies are randomly selected and assigned with magnitude of 1, the rest frequencies are set to 0. Each data point in the result graphs is the average over 100 runs with different images. All experiments were carried out on the system that is equiped with 8 Intel Xeon E7-8830 2.13GHz CPU total of 64 cores and 1TB memory.

\subsection{Runtime}

\subsubsection{Runtime vs. Signal Size}
\begin{figure}[htbp]
\begin{minipage}{0.49\linewidth}
  \centerline{\includegraphics[width=1\textwidth]{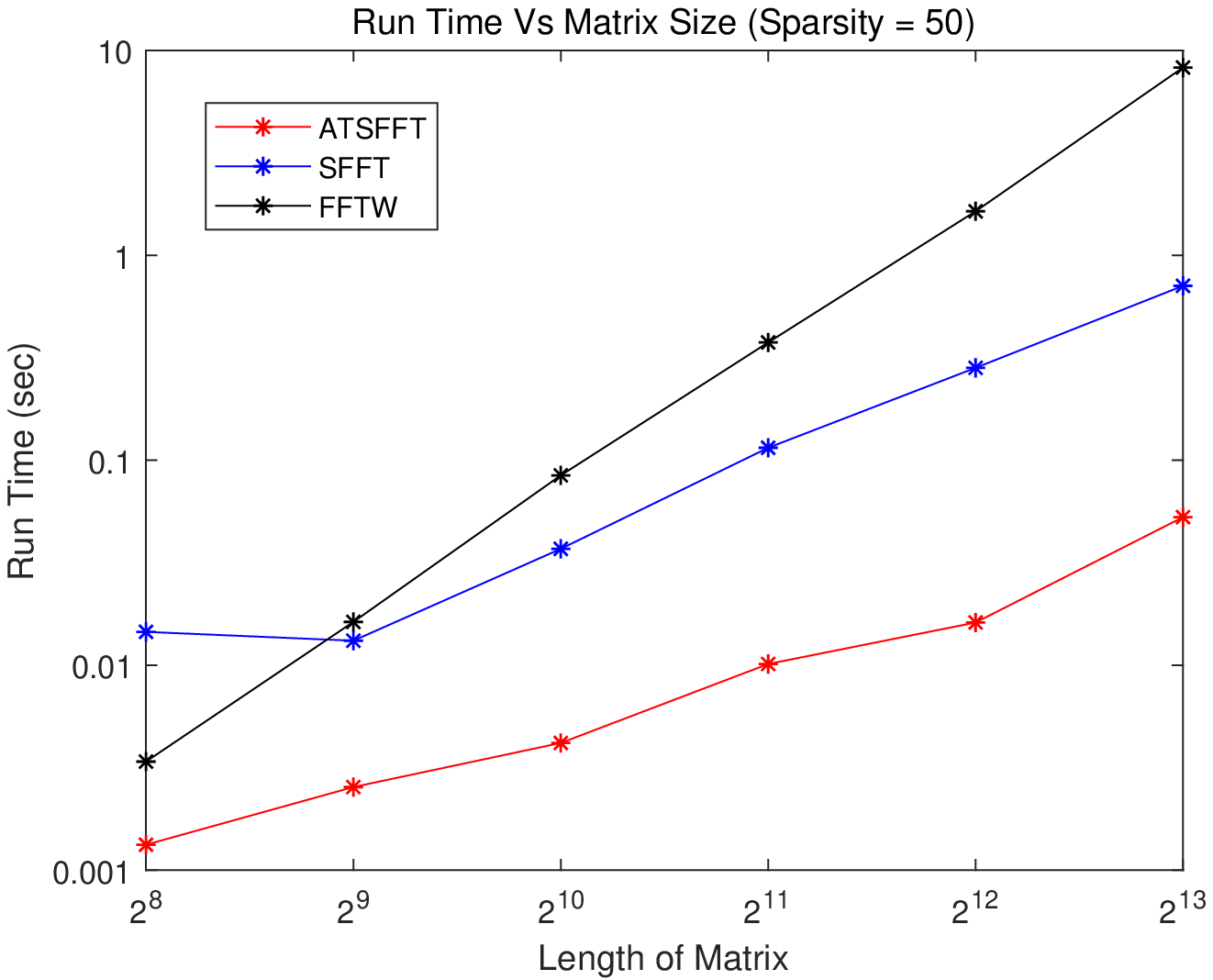}}
  \centerline{\scriptsize{(a)}}
  \centerline{}
\end{minipage}
\hfill
\begin{minipage}{0.49\linewidth}
  \centerline{\includegraphics[width=1\textwidth]{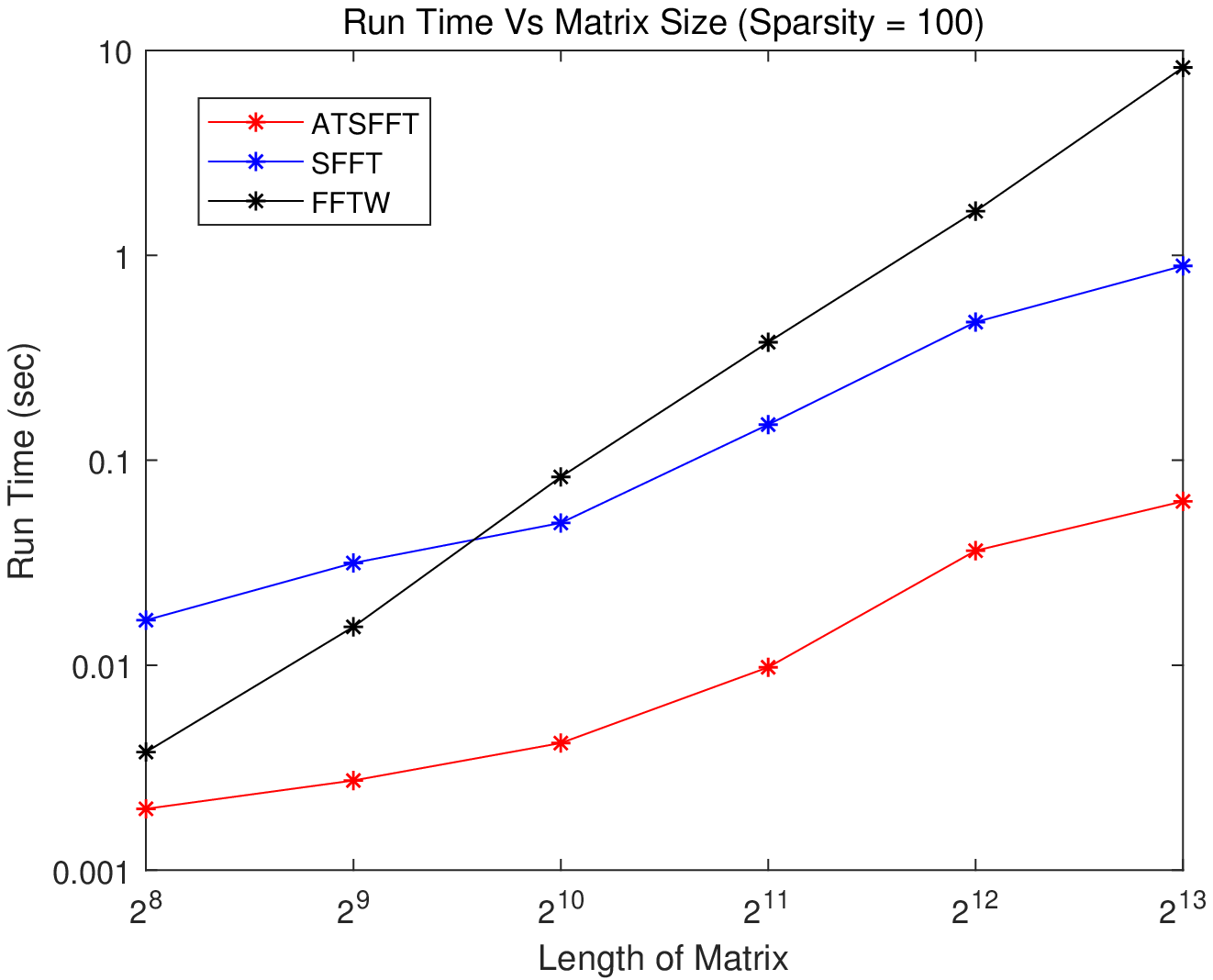}}
  \centerline{\scriptsize{(b)}}
  \centerline{}
\end{minipage}
\caption{The runtime comparison of ATSFFT, SFFT and FFTW on different signal sizes (a) the sparsity parameter is fixed to $(k=50)$,(b) the sparsity parameter is ficed to $(k=100)$}
\label{fig:8}
\end{figure}

The sparsity parameter $k$ is fixed to constant $k=50$ and $k=100$, there are 6 different image matrix size from $2^8\times2^8$ to $2^{13}\times2^{13}$. The average runtime are shown in Figure~\ref{fig:8}. We can see that SFFT runs faster than FFTW in most cases. However, ATSFFT outperforms its competitors by an order of magnitude. It is no surprise to see that the runtime of the three algorithms are approximately linear in the log scale with respect to problem size. Furthermore, the runtime of ATSFFT grows with the smallest slope than the others as the problem size increases.

In Figure~\ref{fig:8}(a), it shows that SFFT is faster than FFTW while recovering the exact 50 non-zero coefficients for bigger problem size, and $N={2^9}$ is the breaking dimension size. When signal size is less than${2^9}\times{2^9}$, SFFT is slower than FFTW. ATSFFT is the fastest among the three algorithms for all cases.  Similarly, Figure~\ref{fig:8}(b) shows that for signal dimension size $N>{2^{10}}$ SFFT is faster than FFTW while recovering the exact 100 non-zero coefficients. When signal size is less than ${2^{10}}\times{2^{10}}$, SFFT is slower than FFTW. Again, ATSFFT is the fastest for all cases. The results show that the ATSFFT significantly extends the range of applications of sparse FFT where sparse approximation is applicable.
\begin{figure}[htbp]
\begin{minipage}{0.49\linewidth}
  \centerline{\includegraphics[width=1\textwidth]{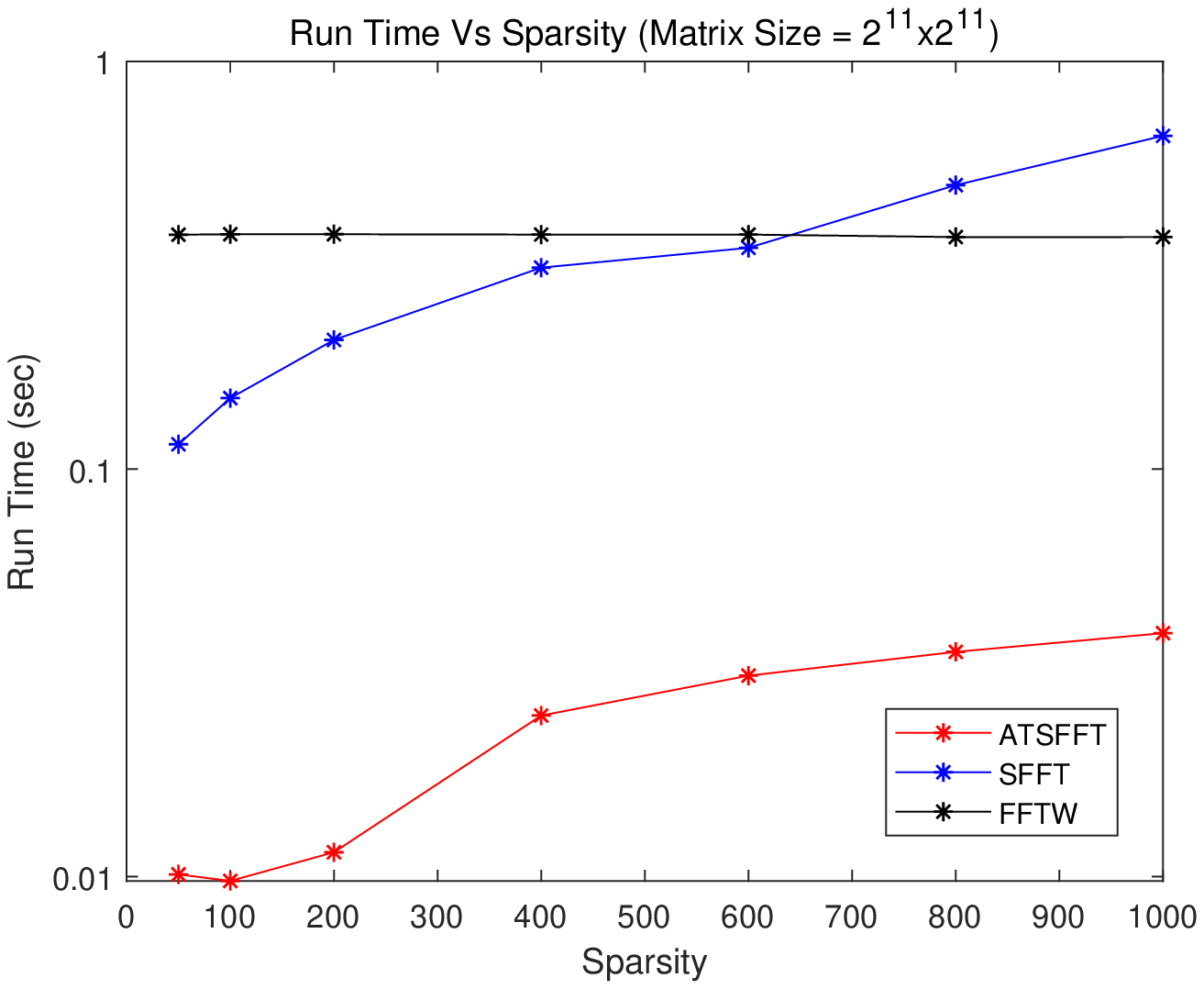}}
  \centerline{\scriptsize{(a)}}
  \centerline{}
\end{minipage}
\hfill
\begin{minipage}{0.49\linewidth}
  \centerline{\includegraphics[width=1\textwidth]{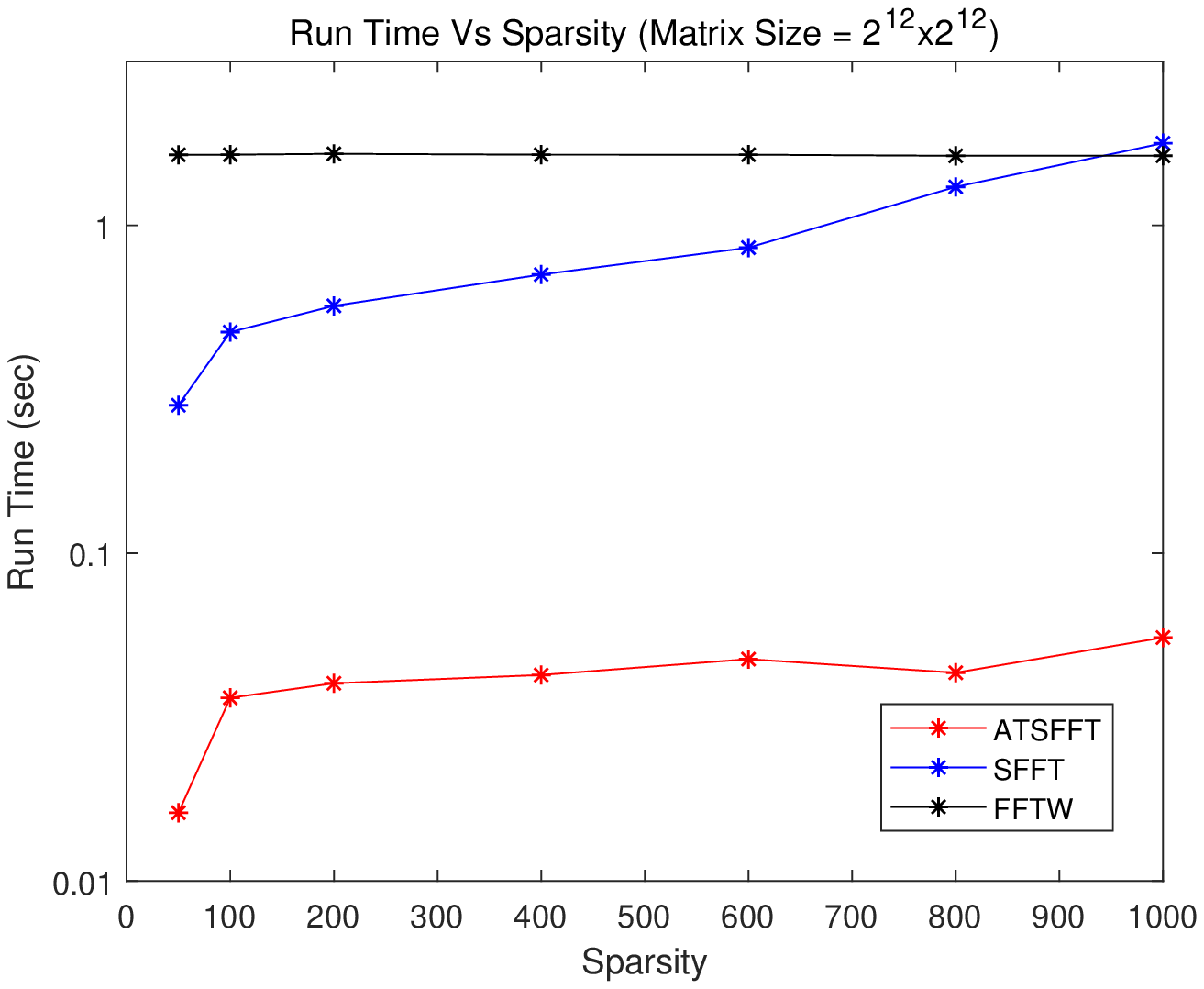}}
  \centerline{\scriptsize{(b)}}
  \centerline{}
\end{minipage}
\caption{The runtime comparison of ATSFFT, SFFT and FFTW on different signal sparsity (a) the image matrix size is fixed to $2^{11}\times2^{11}$,(b) the image matrix size is fixed to $2^{12}\times2^{12}$}
\label{fig:9}
\end{figure}
\subsubsection{Runtime vs. Number of Non-zero Frequency}
The image matrix size is fixed to constant ${2^{11}}\times{2^{11}}$ and ${2^{12}}\times{2^{12}}$  and the runtime of the comparing algorithms for sparsity $k (k=50, 100, 200, 400, 600, 800, 1000)$ are shown in Figure~\ref{fig:9}. For each value of $k$, the experiment is repeated 100 times. As the sparsity $k$ increases, ATSFFT and SFFT take more time to complete, while the runtime of FFTW which depends on image size but not sparsity $k$ is essentially constant.

Figure~\ref{fig:9}(a) shows that SFFT runs faster than FFTW when sparsity $k<700$, and SFFT presents disadvantage when sparsity $k>=800$. While the runtime of ATSFFT for all cases keeps under 0.1 second, ATSFFT outperforms SFFT and FFTW.  Similarly, Figure~\ref{fig:9}(b) shows SFFT is faster than FFTW when sparsity $k<900$, and runs slower otherwise. ATSFFT appears to be the fastest where the runtimes for all cases are under 0.1 seconds as well. The above experimental results show that ATSFFT significantly extends the range of applications for which sparse approximation of SFFT is practical.

\subsection{Speedup}
An important question we would like to address is that, performance wise,  how does ATSFFT compare with SFFT and FFT.

\subsubsection{Speedup vs. Signal Size}
\begin{table*}[htbp]
  \centering
  \caption{The speedup of ASFFT and SFFT over FFTW with fixed sparsity}
    \begin{tabular}{|c|c|c|c|c|c|c|}
    \hline
    \multicolumn{1}{|c|}{\textbf{$k=50$}}&\multicolumn{6}{c|}{speedup}\bigstrut\\
    \hline
    N   & $2^8$  & $2^9$  & $2^{10}$  & $2^{11}$  &  $2^{12}$  & $2^{13}$   \bigstrut\\
    \hline
    ATSFFT/SFFT & $10.9\times$  & $5.2\times$  & $8.9\times$ & $11.4\times$ & $17.5\times$ & $13.5\times$ \bigstrut\\
    \hline
    SFFT/FFTW  & $0.2\times$  & $1.2\times$  & $2.3\times$  &  $3.3\times$ & $5.8\times$  &$11.6\times$\bigstrut\\
    \hline
    \multicolumn{1}{|c|}{\textbf{$k=100$}}&\multicolumn{6}{c|}{speedup}\bigstrut\\
    \hline
    N   & $2^8$  & $2^9$  & $2^{10}$  & $2^{11}$  &  $2^{12}$  & $2^{13}$   \bigstrut\\
    \hline
    ATSFFT/SFFT & $8.3\times$  & $11.5\times$  & $11.9\times$ & $15.3\times$ & $13.1\times$ & $14.1\times$ \bigstrut\\
    \hline
    SFFT/FFTW  & $0.2\times$  & $0.5\times$  & $1.7\times$  &  $2.5\times$ & $3.5\times$  &$9.3\times$\bigstrut\\
    \hline
    \end{tabular}%
  \label{tab:1}%
\end{table*}
Table~\ref{tab:1} shows the speedup of ATSFFT to SFFT and SFFT to FFTW where the sparsity is set to be $(k=50)$ and $(k=100)$. For sparsity $(k=50)$,  except that SFFT runs slower than FFTW for the case of $N={2^8}$, SSFT achieves more speedup when image size increases. However,  ATSFFT achieves better performance across the board, and by average a magnitude of speedup than SFFT. For sparsity $(k=100)$,  similar observation is obtained.  In conclusion, ATSFFT exhibits strong performance and stability over its peers.

\subsubsection{Speedup vs. Number of Non-zero Frequency}
\begin{table*}[htbp]
  \centering
  \caption{The speedup of ASFFT, SFFT and FFTW with fixed image size}
    \begin{tabular}{|c|c|c|c|c|c|c|c|}
    \hline
    \multicolumn{1}{|c|}{\textbf{$N=2^{11}$}}&\multicolumn{7}{c|}{Speedup}\bigstrut\\
    \hline
    $k$   & $50$  & $100$  & $200$  & $400$  &  $600$  & $800$  & $1000$ \bigstrut\\
    \hline
    ATSFFT/SFFT & $11.4\times$	& $15.3\times$ & $18.1\times$	& $12.6\times$ & $11.2\times$  & $14.0\times$ &$16.6\times$  \bigstrut\\
    \hline
    SFFT/FFTW  & $3.3\times$	& $2.5\times$ &	$1.8\times$ &	$1.2\times$	& $1.1\times$ &	$0.8\times$  &   $0.6\times$\bigstrut\\
    \hline
    \multicolumn{1}{|c|}{\textbf{$N=2^{12}$}}&\multicolumn{7}{c|}{Speedup}\bigstrut\\
    \hline
    $k$   & $50$  & $100$  & $200$  & $400$  &  $600$  & $800$  & $1000$ \bigstrut\\
    \hline
    ATSFFT/SFFT & $17.5\times$	& $13.1\times$ & $14.2\times$	& $16.7\times$ & $18.0\times$  & $30.3\times$ &$32.2\times$  \bigstrut\\
    \hline
    SFFT/FFTW  & $5.8\times$	& $3.5\times$ &	$2.9\times$ &	$2.3\times$	& $1.9\times$ &	$1.2\times$  &   $0.9\times$\bigstrut\\
    \hline
    \end{tabular}%
  \label{tab:2}%
\end{table*}
In this section, we compare the performance with regard to different number of non-zero frequencies when the input image size is fixed. Table~\ref{tab:2} shows the speedup of ATSFFT to SFFT and SFFT to FFTW for image matrix size of ${2^{11}}\times{2^{11}}$ and ${2^{12}}\times{2^{12}}$.  For both image sizes, we can see that SFFT performs the best when the sparsity is the smallest, and the speedup over FFTW reduces when the sparsity number $K$ increasing. However, ATSFFT could maintain performance superiority more consistently. For image size ${2^{12}}\times{2^{12}}$, when sparsity is 1000, ATSFFT is about 29 times faster than FFTW while SFFT runs slightly slower than FFTW. Again, the ATSFFT demonstrates robustness over sparsity variation.

\subsection{Error}

We compute the error metric per as the average $L_1$ error between the output $k-$sparse approximation $\hat x^{'}$ and the fft of $x$ referred to as $\hat{x}$.

\begin{equation}
Error={\frac{1}{k}}\sum\limits_{i=0}^{N-1}\sum\limits_{j=0}^{N-1}|\hat x^{'}_{i,j}-\hat x_{i,j}|\\
\end{equation}

\begin{table*}[htbp]
  \centering
  \caption{The Error of ASFFT and SFFT}
    \begin{tabular}{|c|c|c|c|c|c|c|}
    \hline
    \multicolumn{1}{|c|}{\textbf{$k=50$}}&\multicolumn{6}{c|}{Error}\bigstrut\\
    \hline
    N   & $2^8$  & $2^9$  & $2^{10}$  & $2^{11}$  &  $2^{12}$  & $2^{13}$   \bigstrut\\
    \hline
    ATSFFT & $5.35\times10^{-03}$  & $3.15\times10^{-05}$  & $1.92\times10^{-05}$ & $5.17\times10^{-10}$ & $5.38\times10^{-10}$ & $1.33\times10^{-10}$ \bigstrut\\
    \hline
    SFFT  & $8.54\times10^{-03}$  & $2.27\times10^{-04}$  & $7.19\times10^{-05}$  &  $5.58\times10^{-10}$ & $1.44\times10^{-09}$  &$7.46\times10^{-11}$\bigstrut\\
    \hline
    \multicolumn{1}{|c|}{\textbf{$k=100$}}&\multicolumn{6}{c|}{Error}\bigstrut\\
    \hline
    N   & $2^8$  & $2^9$  & $2^{10}$  & $2^{11}$  &  $2^{12}$  & $2^{13}$   \bigstrut\\
    \hline
    ATSFFT & $3.66\times10^{-04}$  & $2.10\times10^{-04}$  & $3.08\times10^{-07}$ & $2.08\times10^{-07}$ & $3.69\times10^{-10}$ & $2.02\times10^{-09}$ \bigstrut\\
    \hline
    SFFT  & $5.46\times10^{-04}$  & $2.84\times10^{-04}$  & $6.36\times10^{-06}$  &  $7.94\times10^{-07}$ & $2.38\times10^{-10}$  &$2.32\times10^{-10}$\bigstrut\\
    \hline
    \end{tabular}%
  \label{tab:3}%
\end{table*}
\subsubsection{Error vs. Signal Size}

Tab.~\ref{tab:3} shows the error of the compared algorithms ATSFFT and SFFT where the sparsity parameter $k$ is fixed to constant $(k=50)$ and $(k=100)$ and the image matrix size is from $2^8\times2^8$ to $2^{13}\times2^{13}$. Smaller is better. For sparsity $(k=50)$, the error of ATSFFT and SFFT decreased with the image size increasing. The error of ATSFFT is smaller than SFFT. Similar results appear in the experiments for sparsity $(k=100)$. This shows the sparser the image signal, the smaller the error of ATSFFT and SFFT. Morever, the error of ATSFFT is smaller than SFFT, which indicate ATSFFT can control the error better than SFFT.

\subsubsection{Error vs. Number of Non-zero Frequency}

In this section, we compare the error with regard to different number of non-zero frequencies and different image size. Tab.~\ref{tab:4} shows the error of ATSFFT and SFFT increased with the image sparsity increasing. The error of ATSFFT is smaller than SFFT. We can draw the conclusion the sparser the image signal, the smaller the error of ATSFFT and SFFT. Morever, ATSFFT can control the error better than SFFT.

\begin{table*}[htbp]
  \centering
  \caption{The Error of ASFFT, SFFT and FFTW}
    \begin{tabular}{|c|c|c|c|c|c|c|c|}
    \hline
    \multicolumn{1}{|c|}{\textbf{$N=2^{11}$}}&\multicolumn{7}{c|}{Error}\bigstrut\\
    \hline
    N   & $50$  & $100$  & $200$  & $400$  &  $600$  & $800$  & $1000$ \bigstrut\\
    \hline
    ATSFFT & $5.17\times10^{-10}$	& $2.08\times10^{-07}$ & $3.43\times10^{-05}$	& $9.23\times10^{-07}$ & $8.02\times10^{-06}$  & $1.45\times10^{-05}$ & $1.65\times10^{-05}
$  \bigstrut\\
    \hline
    SFFT  & $5.58\times10^{-10}$	& $7.94\times10^{-07}$ &	$8.72\times10^{-05}$ &	$1.49\times10^{-06}$	& $2.57\times10^{-05}$ &	$3.70\times10^{-05}$  &   $3.18\times10^{-05}$\bigstrut\\
    \hline
    \multicolumn{1}{|c|}{\textbf{$N=2^{12}$}}&\multicolumn{7}{c|}{Error}\bigstrut\\
    \hline
    N   & $50$  & $100$  & $200$  & $400$  &  $600$  & $800$  & $1000$ \bigstrut\\
    \hline
    ATSFFT & $5.38\times10^{-10}$	& $3.69\times10^{-10}$ & $1.44\times10^{-09}$	& $2.54\times10^{-05}$ & $4.73\times10^{-05}$  & $1.63\times10^{-05}$ & $1.77\times10^{-05}
$  \bigstrut\\
    \hline
    SFFT  & $1.44\times10^{-9}$	& $2.38\times10^{-10}$ &	$1.60\times10^{-09}$ &	$2.64\times10^{-06}$	& $1.27\times10^{-05}$ &	$1.87\times10^{-05}$  &   $2.25\times10^{-05}$\bigstrut\\
    \hline
    \end{tabular}%
  \label{tab:4}%
\end{table*}

\section{Conclusion}
\label{sec:conclusion}

The Sparse Fast Fourier Transform (SFFT) is a novel algorithm for discrete Fourier transforms on signals with the sparsity in frequency domain. However, the SFFT implementation has the drawback that it only works reliably for very specific input parameters, especially signal sparsity $k$. This drawback hinders the extensive applications of SFFT. we propose an Adaptive Tuning Sparse Fast Fourier Transform (ATSFFT), which is a novel sparse fast fourier transform enabled with sparsity detection. In the case of unknown sparsity $k$, ATSFFT can probe the sparsity $k$ via adaptive dynamic tuning technology and then complete the Fourier transform of signal. We present some numerical experiments comparing runtime, speedup and error of ATSFFT, SFFT and FFT in FFTW algorithms library. Experimental results show that ATSFFT not only can control the error better than SFFT, but also performs faster than SFFT, which computes more efficiently than the state-of-the-art FFTW.


\newpage


\begin{thebibliography}{9}

\bibitem{jour0} C.F.Beckmann and S.M.Smith, "Probabilistic independent component analysis for functional magnetic resonance imaging," IEEE Transactions on Medical Imaging, vol.23,no.2,pp.137-152,2004.

\bibitem{jour01} Shota Taki, Fumihiko Sakaue, and Jun Sato, "High resolution light field photography from split ray imaging and coded aperture," in VISAPP 2014-Proceedings of the 9th International Conference on Computer Vision Theory and Applications, Volume 2, Lisbon, Portugal, 5-8 January, 2014, pp.605-612.

\bibitem{jour02} Rik Jongerius, Strfan J. Wihnholds. Ronald Nijboer, and Henk Corporal, "An end-to-end computing model for the square kilometre array," IEEE Computer, vol. 47, no.9,pp.48-54, 2014.

\bibitem{jour03} Jon Atli Benediktsson, Martino Pesaresi, and Kolbeinn Amason, "Classsfication and feature extraction for remote sensing images from urban areas based on morphological transformations," IEEE Trans. Geoscience and Remote Sensing, vol. 41, no.9,pp.1940-1949, 2003.

\bibitem{jour04} S. Grace Chang, Bin Yu, and Martin Vetterli, "Adaptive wavelet thresholding for image denoising and compression," IEEE Trans. Image Processing, vol.0, no.9,pp.1532-1546, 2000.


\bibitem{jour1} X. Zhang, R. Xiong, W. Lin, S. Ma, J. Liu and W. Gao, "Video Compression Artifact Reduction via Spatio-Temporal Multi-Hypothesis Prediction," in IEEE Transactions on Image Processing, vol. 24, no. 12, pp. 6048-6061, Dec. 2015.

\bibitem{jour2} B. S. Reddy and B. N. Chatterji, "An FFT-based technique for translation, rotation, and scale-invariant image registration," in IEEE Transactions on Image Processing, vol. 5, no. 8, pp. 1266-1271, Aug 1996.

\bibitem{jour3} X. Zhang; R. Xiong; W. Lin; J. Zhang; S. Wang; S. Ma; W. Gao, "Low-Rank based Nonlocal Adaptive Loop Filter for High Efficiency Video Compression," in IEEE Transactions on Circuits and Systems for Video Technology , vol.27,no.10, pp.2177-2188, Oct. 2017.

\bibitem{jour4} S. Ma, X. Zhang, J. Zhang, C. Jia, S. Wang and W. Gao, "Nonlocal In-Loop Filter: The Way Toward Next-Generation Video Coding?," in IEEE MultiMedia, vol. 23, no. 2, pp. 16-26, Apr.-June 2016.

\bibitem{jour5} X. Zhang, W. Lin, R. Xiong, X. Liu, S. Ma and W. Gao, "Low-Rank Decomposition-Based Restoration of Compressed Images via Adaptive Noise Estimation," in IEEE Transactions on Image Processing, vol. 25, no. 9, pp. 4158-4171, Sept. 2016.

\bibitem{book1} X. Zhang, W. Lin, S. Wang and S. Ma, "Nonlocal Adaptive In-Loop Filter via Content-Dependent Soft-Thresholding for HEVC," 2015 IEEE International Symposium on Multimedia (ISM), Miami, FL, 2015, pp. 465-470.

\bibitem{book2} X. Zhang, W. Lin, J. Liu and S. Ma, "Compression noise estimation and reduction via patch clustering," 2015 Asia-Pacific Signal and Information Processing Association Annual Summit and Conference (APSIPA), Hong Kong, 2015, pp. 715-718.

\bibitem{jour6} X. Zhang, R. Xiong, X. Fan, S. Ma and W. Gao, "Compression Artifact Reduction by Overlapped-Block Transform Coefficient Estimation With Block Similarity," in IEEE Transactions on Image Processing, vol. 22, no. 12, pp. 4613-4626, Dec. 2013.

\bibitem{book3}S. Shi, R. Xiong, S. Ma, X. Fan and W. Gao, "Image compressive sensing using overlapped block projection and reconstruction," 2015 IEEE International Symposium on Circuits and Systems (ISCAS), Lisbon, 2015, pp. 1670-1673.

\bibitem{book4} Haitham Hassanieh, Piotr Indyk, Dina Katabi and Eric Price,"Simple and practical algorithm for sparse Fourier transform," Proceedings of the Twenty-Third Annual {ACM-SIAM} Symposium on Discrete Algorithms, 2012, Japan, pp. 1183--1194.

\bibitem{book5} Haitham Hassanieh, Piotr Indyk, Dina Katabi and Eric Price,"Nearly optimal sparse fourier transform," Proceedings of the 44th Symposium on Theory of Computing Conference, 2012, New York, pp. 563--578.

\bibitem{book6}S. Shi, R. Yang and H. You, "A new two-dimensional Fourier transform algorithm based on image sparsity," 2017 IEEE International Conference on Acoustics, Speech and Signal Processing (ICASSP), New Orleans, LA, 2017, pp. 1373-1377.

\bibitem{book7}S. Shi, R. Yang, X. Zhang, H. You, D. Fan, "An Adaptive Tuning Sparse Fast Fourier Transform", Advances in Multimedia Information Processing (PCM) 2017. Lecture Notes in Computer Science, vol 10736.

\bibitem{book8}M. Frigo and S. G. Johnson, "FFTW: an adaptive software architecture for the FFT," Acoustics, Speech and Signal Processing, 1998. Proceedings of the 1998 IEEE International Conference on, Seattle, WA, 1998, pp. 1381-1384.

\end{thebibliography}
\end{document}